\documentclass[%
,aps%
 ,onecolumn%
 ,secnumarabic%
,amssymb, amsmath, showkeys, nofootinbib, floatfix]{revtex4-1}
\usepackage{tikz}
\usepackage{graphicx}
\usepackage{hyperref}
\usepackage{siunitx}
\usepackage{amsmath}
\usepackage{bm}
\usepackage{color}
\usepackage{url}
\usepackage[american]{babel}
\newcommand{\extended}[1]{{\color{black} #1}}
\def \hpftouC {\num{7E-5}}
\def \hpftouD {\num{3E-4}}
\def \hpftouE {\num{3E-3}}
\DeclareSIUnit\square{sq}
\newcommand\copyrighttext{%
  \footnotesize \textcopyright 2016 IEEE. Personal use of this material is permitted.
  Permission from IEEE must be obtained for all other uses, in any current or future 
  media, including reprinting/republishing this material for advertising or promotional 
  purposes, creating new collective works, for resale or redistribution to servers or 
  lists, or reuse of any copyrighted component of this work in other works. 
  DOI: \href{http://dx.doi.org/10.1109/TIM.2016.2570127}{10.1109/TIM.2016.2570127}}
\newcommand\copyrightnotice{%
\begin{tikzpicture}[remember picture,overlay]
\node[anchor=south,yshift=10pt] at (current page.south) {\fbox{\parbox{\dimexpr\textwidth-\fboxsep-\fboxrule\relax}{\copyrighttext}}};
\end{tikzpicture}%
}

\begin{document}
\title{Electrical Resistance Tomography of Conductive Thin Films}%
\author{Alessandro \surname{Cultrera}}%
\email{a.cultrera@inrim.it}%
\author{Luca \surname{Callegaro}}%
\affiliation{INRIM - Istituto Nazionale di Ricerca Metrologica, strada delle Cacce, 91, 10035 Torino, Italy.}%
\date{June 2016}%


%
\copyrightnotice
\begin{abstract}
The Electrical Resistance Tomography (ERT) technique is applied to the measurement of sheet conductance maps of both uniform and patterned conductive thin films. Images of the sheet conductance spatial distribution, and local conductivity values are obtained.
Test samples are tin oxide films on glass substrates, with electrical contacts on the sample boundary; some samples are deliberately patterned in order to induce null conductivity zones of known geometry while others contain higher conductivity inclusions. Four-terminal resistance measurements among the contacts are performed with a scanning setup.
The ERT reconstruction is performed by a numerical algorithm based on the total variation regularization and the L-curve method. 
ERT correctly images the sheet conductance spatial distribution of the samples.
The reconstructed conductance values are in good quantitative agreement with independent measurements performed with the van der Pauw and the four-point probe methods. 
\end{abstract}

\maketitle

\section{introduction}
Electrical Impedance Tomography (EIT)~\cite{cheney1999eit, borcea2002electrical, seo2013electrical} is a technique that allows the non-invasive measurement of spatial distribution of the electrical properties of an object, by performing electrical measurements on its boundary. 
These electrical properties can be related to local chemical composition or other physical properties. 
EIT founds typical applications in clinical, earth and civil sciences. Some examples are in the imaging of pulmonary ventilation, of hydric reservoirs, in geological site monitoring and process engineering~\cite{murray2015use, revil2012review, zhang2010identification}. 

In the dc regime EIT is called Electrical Resistance Tomography (ERT); it reconstructs the resistivity/conductivity spatial distributions within an object. 
As a fundamental advantage ERT represents a non-scanning technique that can retrieve local information. This allows to work on relatively large area samples and require a quite simple measurement setup compared to other local techniques like those based on scanning probes.

In this work, we present an application of ERT to the measurement of the conductivity spatial distribution of conductive thin film samples, to map the topography of their defects, inhomogeneities, and inclusions. Up to now the only earlier similar application the authors found in literature dealt with doped silicon wafers, but they recovered equally doped profiles rather than a conductivity map~\cite{djamdji1996electrical}. 
The application is based on a measurement setup realised with commercial instrumentation and open-source data analysis software. 

The samples here investigated to test the technique are commercial tin-oxide films on glass substrate, some of which have been patterned with a scriber. The induced conductivity defects, having known position and shape, are compared with the ERT conductivity reconstructions. The results are further validated with van der Pauw (vdP) measurements~\cite{van1958method} on uniform samples, performed with the same experimental setup, and by means of four-point probe measurements on both types of sample. 

\section{Background \label{background}}
An ERT experiment involves the connection to the sample of $N$ contacts, and the measurement of four-terminal resistances among these contacts. A typical measurement protocol involves the injection of a current between two adjacent contacts, and the measurement of the potential differences $\bm{V}_\textup{m}$ between the remaining available contacts,\footnote{Two- or three-terminal transresistances are typically not considered because they include the unknown contact resistance.} resulting in a list of $N(N-3)$ transresistances. Other protocols, combining the contacts in different order, exist.

Let us assume that the sample is two-dimensional, isotropic and linear, and can thus be described by a continuous conductivity distribution $\bm{\sigma(\bm{r})}$, being $\bm{r}$ the position vector. 
The reconstruction of $\bm{\sigma(\bm{r})}$ from the discrete list of measured resistances is an ill-posed inverse problem, because from the bare knowledge of data on the boundary a a solution on a two-dimensional domain is searched. The reconstruction is sensitive to noise in the input data, and some prior knowledge on $\bm{\sigma(\bm{r})}$ (e.g., smoothness, piece-wise continuity) is needed. So-called \emph{regularization} methods are required. Several \extended{books and} papers in literature focused on the way to attack such problem in the most effective way (see~\cite{tikhonov2013numerical} for an overview, and~\cite{yorkey1987comparing, borsic2007total, jung2015impedance}); our approach is discussed in paragraphs~\ref{section:code} and~\ref{section:lcurve}.

The reconstruction process consists in the minimization of a functional in $\bm{\sigma}$, which contains the voltages $\bm{V}_\textup{m}$ measured at domain boundary, and the \emph{forward model} $\bm{U}(\bm{\sigma})$, the set of calculated voltages $\bm{U}$ for a given $\bm{\sigma}$ :

\begin{equation}
\begin{aligned}
\label{eqn:tikhonov}
&\underset{\bm{\sigma}}{\textup{min}} 
& \left \{ ||\bm{V}_\textup{m}-\bm{U}(\bm{\sigma})||^{2}+\alpha^{2} \mathcal{R}(\bm{\sigma})\right \}.
\end{aligned}
\end{equation}

The \emph{residual term} $||\bm{V}_\textup{m}-\bm{U}(\bm{\sigma})||$ is the euclidean distance between $\bm{V}_\textup{m}$ and $\bm{U}(\bm{\sigma})$. The \emph{regularization term} $\mathcal{R}(\bm{\sigma})$ includes the constraining information on $\bm{\sigma}$. The tuning constant $\alpha$ is called the regularization parameter, or \emph{hyper-parameter}. Several forms for the  regularization term have been proposed~\cite{holder2004electrical}.
The solution of~\eqref{eqn:tikhonov} is carried out by numerical analysis, typically with finite element methods (see paragraph~\ref{section:code}). Depending on the choice of the regularization term, it may require the iterative calculation of the forward model, thus being computationally intensive.
The choice of $\alpha$ is not trivial. A too large $\alpha$ over-smooths the result whereas a too small $\alpha$ introduces noise-related artefacts. In our application we use the so-called L-curve method (see paragraph~\ref{section:lcurve}).  

Heuristic methods can be acceptable in clinical application since the expected morphology of the observed system (human body) is a straightforward assumption~\cite{graham2006objective}, while more restrictive approach is mandatory in materials science, where the topography of a sample may be very different from expected or desired. At an earlier stage of the present experiment we set the amount of regularization $\alpha$ by an empirical method based on the boundary voltage measurements accuracy. 

The present work constitutes an extension of a previous contribution~\cite{RTSI}; here we expand the discussion on the objective selection of the correct amount of $\alpha$, an analysis of the measurement setup performance, and an expression of the spatial resolution of the reconstructed maps. Moreover we investigated also highly conductivity defects not discussed in the previous work.

\section{Experimental\label{section:experimental}}
\begin{figure}[t]
\centering
\includegraphics[width=3.5 in]{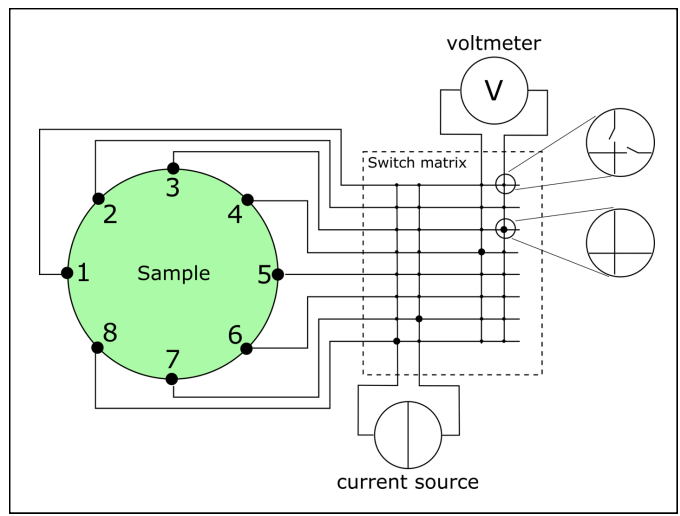}
\caption{Experimental setup scheme. The power supply and digital multimeter modules drive current and measure voltage drops through the relay matrix module (dashed rectangle). The latter allows four-wire measurement combination consistent with a given ERT protocol. As an example, thicker dots represent one possible configuration of relay switches (see circular insets for open/closed notation) connecting rows and columns of the matrix that corresponds to driving current through the sample at electrode pair (7,8) while measuring the voltage drop at electrode pair (3,4).}
\label{fig:expsetup}
\end{figure}	

\subsection{Measurement setup}
The measurements are performed with the setup shown in Fig.~\ref{fig:expsetup}. The four-terminal resistances are measured with the $I$-$V$ method, by using a dc current source, a voltmeter and a $N\times 4$ relay switching matrix.\footnote{Keysight 34980A equipped with: Keysight 34951 4-channel D/A converter, the internal voltmeter of the 34980A and a Keysight 34933 reed matrix.} The switching matrix is programmed to perform the stimulation/measurement patter consistent with the adjacent protocol described in section~\ref{background}. Data acquisition is handled by means of direct \texttt{MATLAB}\texttrademark-instrument communication via IEEE-488 bus. An ERT measurement cycles takes about \numrange{5}{15} minutes depending on the voltmeter accuracy settings.
The same setup can be used to perform van der Pauw resistivity measurements\cite{van1958method}.
As a crosscheck, resistivity measurements can be performed with a commercial four-point probe head\footnote{Jandel cylindrical four-point probe head, \SI{0.635}{\milli\meter} needle spacing.} connected to a  source-meter.\footnote{Keithley 2410 source-meter.} 	
	
\subsection{Samples}
\label{sec:samples}
Our samples consisted of commercial Fluorine-doped Tin Oxide films (FTO,  SnO$_{2}$:F) on glass substrates. FTO is highly doped and its electrical behaviour is ohmic\cite{cultrera2014band}. The typical application of FTO is in the manufacturing of touch screens and dye-sensitized solar cells~\cite{gratzel:2005}. 
The samples chosen\footnote{Solaronix TEC 7.} have a film thickness of about \SI{500}{\nano\metre} and a nominal sheet resistance of \SI{7}{\ohm\per\square} corresponding to \SI{66}{\milli\siemens\square}.\footnote{We remark that even though factory specifications are expressed in units of sheet resistance, in the following the reference unit would be sheet conductance. This because resistance is not appropriate to treat the  presence of features of infinite resistance. The authors used sometimes the term ``conductivity'' in place of ``sheet conductance'' when discussing mathematical or qualitative aspects.}  
A laser scriber was used to define accurate samples and contacts geometry. The diameter of the sample area was \SI{10}{mm}.\footnote{Harrison Laser Pro Mercury \SI{12}{\watt}.} 
 Eight equally spaced pads were defined at the same time at sample boundary, being actually \textit{in-situ} electrodes, to host measurement leads, see Fig.~\ref{fig:fto2u}-a and -b, Fig.~\ref{fig:large}-a and -b. Conductive silver paste was used to glue the measurement leads from the sample to the measurement setup.
Three types of samples have been realised: Uniform samples (labeled FTOU in the following), see Fig.~\ref{fig:fto2u}-a, where the inner area of the sample is continuous; and non-uniform, Defective samples (FTOD), see Fig.~\ref{fig:large}-a, where non conductive zones inside the sample area were cut out, and samples with Highly conductive defects (FTOH), see Fig.~\ref{fig:ftoAg}-a. The latter were obtained placing a drop of colloidal silver paste on a uniform sample.\footnote{RS Silver conductive paste 186-3600.}

\subsection{Code\label{section:code}}
The reconstruction software is \texttt{EIDORS}~v. 3.7.1)\cite{adler2005eidors, adler2006uses}, an open-source function library for \texttt{MATLAB}\texttrademark~(v. R2010b). \texttt{EIDORS} provides a set of finite element modelling tools and allows to choose among different forms for the regularization term $\mathcal{R}(\bm{\sigma})$ in~\eqref{eqn:tikhonov}.
In this work we employ the Total Variation (TV) regularization~\cite{borsic2007total, vauhkonen1998tikhonov, dobson1994image}, allowing a ``blocky'' conductivity prior information and therefore indicated for the sharp conductivity profiles of the samples here investigated. The more common ``Laplace'' prior is instead more indicated for the reconstruction of smooth conductivity distributions. 
The TV regularization term has the form

\begin{equation}
\label{eqn:TV}
\mathcal{R}(\bm{\sigma})= \underset{D}{\int} \! |\nabla \bm{\sigma}| \, \textup{d}D.
\end{equation}

The reconstruction algorithm, is based on the so-called ``primal-dual interior-point method''~\cite{borsic2007total, andersen2000efficient} which is available in \texttt{EIDORS}.

\subsection{L-curve method\label{section:lcurve}}
The L-curve consists in a double log plot of the residual term $||\bm{V}_\textup{m}-\bm{U}(\bm{\sigma})||$ versus the regularization term $\mathcal{R}(\bm{\sigma})$ of~\eqref{eqn:tikhonov}; each point of the curve corresponds to a specific value of the hyper-parameter $\alpha$. See for Fig.~\ref{fig:fto2u}-f and Fig.~\ref{fig:large}-f
The L-curve can be used to find a balance between the quality of the fit, hence a small residual term $||\bm{V}_\textup{m}-\bm{U}(\bm{\sigma})||$, and the complexity of $\bm{\sigma}$, i.e.\footnote{The more flat $\bm{\sigma(r)}$, the smaller $|\nabla \bm{\sigma}|$ in~\eqref{eqn:TV}, and therefore the smaller is the integral $\mathcal{R}(\bm{\sigma})$.} a small regularization term $\mathcal{R}(\bm{\sigma})$. 
Hence, the bottom-left corner of the curve corresponds to the optimal regularization in this perspective. 
\section{Results and Discussion}
\subsection{Instrument performance}\label{sec:performance}
The measurements on the samples described in Sec.~\ref{sec:samples} have been performed by setting the current generator at a fixed current of \SI{1.000(6)}{\milli\ampere}.\footnote{All stated uncertainties correspond to a coverage factor $k=1$.} The voltmeter is set to its \SI{100}{\milli\volt} range, with an aperture time of \SI{50}{\milli\second} (\num{1} number of period line cycles (NPLC)), corresponding to a uncertainty of \SI{2.5}{\micro\volt} (1-year). The measurement repeatability, evaluated on several measurement cycles, is about \SI{4}{\micro\volt}. 
For the samples investigated, the voltage readings are in the range \SIrange{300}{900}{\micro\volt}. The transresistance relative uncertainty can therefore be estimated in the range \SIrange{0.5}{1.5}{\percent}, dominated by measurement noise, and that can be improved by increasing the voltmeter measurement time, by averaging repeated measurements or by increasing the measurement current. However, by analyzing repeated measurement and numerically altering the readings, it has been found that the present uncertainty level is not a limitation factor in the accuracy of the conductivity map reconstructions reported in the following of this Section. 
Table~\ref{table:results} reports: 
\begin{itemize}
\item the nominal sheet conductance of the featured defect (where present);
\item the vdP experimental sheet conductance (for the FTOU sample only). The reported value (and standard deviation) is an average of measurements performed with eight different electrode configurations; 
\item the ERT sheet conductance. For FTOU the reported value is an average over the entire reconstructed map of Fig.~\ref{fig:fto2u}-d; for FTOD it is an average over the mesh elements along the dashed line of Fig.~\ref{fig:large}-d; similarly for FTOH in Fig.~\ref{fig:ftoAg}-b;
\item the sheet conductance measured with the four-point probe method. The result is an average of ten repetitions on the spot marked by $\star$ in Fig.~\ref{fig:fto2u}-b.
\item the ERT average sheet conductance and standard deviation of the defect within the nominal defect area (where present).
\end{itemize}
\subsection{Uniform samples}
%

\begin{table}
\renewcommand{\arraystretch}{1.3}
\caption{Sheet conductance of FTO samples in \si{\milli\siemens\square}. Values in brackets are standard deviations.}
\label{table:results}
\centering
\begin{tabular}{lccccc}
\hline 
\bfseries sample & \bfseries nominal defect $\sigma$ & \bfseries vdP  $\sigma$ & \bfseries ERT  $\sigma$	& \bfseries four-point  $\sigma$ &\bfseries defect  $\sigma$ \\
\hline
FTOU & -	& \num{83(1)}	& \num{83(3)}	&\num{79(15)} 	& - \\
FTOD & \num{0}	& -				& \num{78(6)}	&\num{74(13)}	&\num{7(9)}\\
FTOH & \num{>1000}	& -				& \num{84(3)}	&\num{76(12)}	&\num{300(20)}\\
\hline

\end{tabular}
\end{table}

\begin{figure*}
\centering
\includegraphics[width=6in]{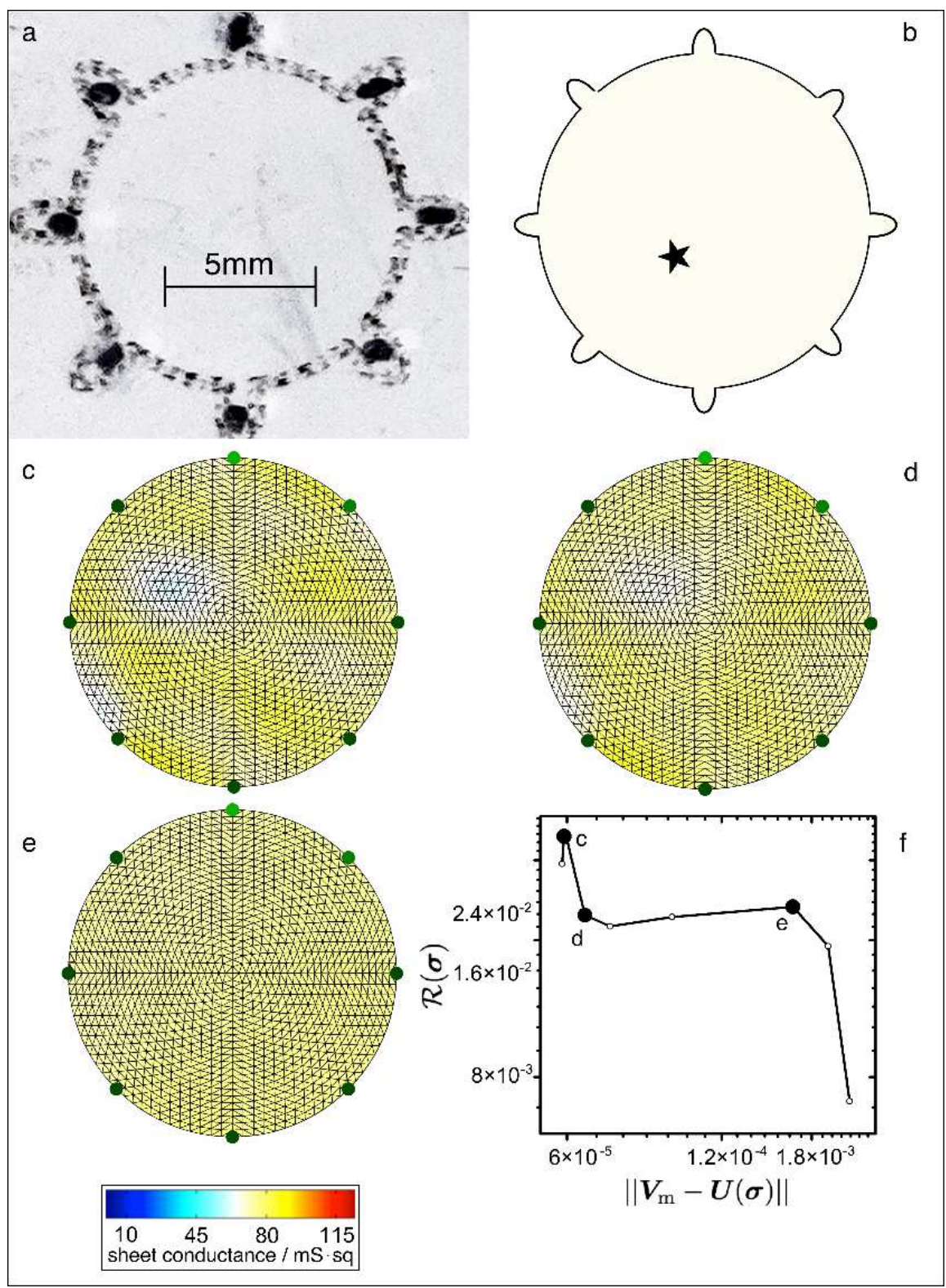}
\caption{FTOU sample. (a) photographic image. (b) schematic diagram: the coulour corresponds to the nominal sheet conductance; the marker $\star$ indicates the four-point probe sheet conductance measurement spot. (c,d,e) sheet conductance maps reconstructed with hyper-parameter values of \hpftouC,~\hpftouD,~\hpftouE~respectively. (d) the L-curve, with markers corresponding to the three different values of $\alpha$. The schematic diagram and the maps follow the same color-scale provided at the bottom.}
\label{fig:fto2u}
\end{figure*}

Fig.~\ref{fig:fto2u}-d shows the reconstructed sheet conductance for the uniform sample FTOU; the hyper-parameter is chosen according to the optimality criterion of section~\ref{section:lcurve} and hence on the left-side corner of the L-curve of Fig.~\ref{fig:fto2u}-f.  
As reported in Table~\ref{table:results}, first row, the experimental values are larger than the nominal one.\footnote{A tentative explanation for the case of FTOU could be that different stocks of FTO plates may present deviations of the sheet conductance from the nominal specifications.} On the other hand, the experimental values agree within the corresponding standard uncertainties. 

\subsection{Non-uniform samples}
Fig.~\ref{fig:large}-d shows the optimal reconstructed sheet conductance map for the sample FTOD. The ERT reconstruction correctly identifies the defective zones of FTOD, both in their position and zero conductivity value since the reconstructed conductivity is $\approx$ \SI{7}{\milli\siemens\square}. 
Fig.~\ref{fig:ftoAg}-b shows the optimal reconstructed sheet conductance map for the sample FTOH. In this case the predicted conductivity of the defect of $\approx$ \SI{300}{\milli\siemens\square} is underestimated. Nevertheless the position of the defect is again correctly located.
As shown in Table~\ref{table:results}, second and third row, the reconstructed sheet conductance value of the undamaged, uniform parts of FTOD and FTOH matches the four-point probe measurements within the corresponding standard deviations (and also very close to the FTOU experimental values).

\subsection{Image reconstruction}
\subsubsection {Spatial resolution}
The definition of proper figures of merit for the quality of an EIT reconstruction is matter of scientific debate. Several metrics have been proposed, mostly related to the use of EIT as an imaging technique, hence quantifying the image resolution, contrast, and geometrical accuracy~\cite{adler2009greit}; these are dependent on the number of electrodes, the measurement stimulation pattern and reading noise, the spatial position and shape of the object to be identified, and the specific reconstruction algorithm employed. 

Here we consider an analogue of the point spread function (PSF) of optical systems. The PSF is roughly assumed to be Gaussian, and is evaluated on both simulated and experimental reconstructions from the edge spread of defects of known geometry. The result is a PSF with a standard deviation (relative to the image diameter) of \SI{5.8}{\percent} for \num{8} electrodes, \SI{3.8}{\percent} for \num{16} electrodes, \SI{3.0}{\percent} for \num{24} electrodes. Increasing the number of electrodes above \num{24} does not reduce the PSF standard deviation with the particular reconstruction algorithm here considered, a conclusion in agreement with other investigations~\cite{Tang2002,Ye2013}.
\begin{figure*}
\centering
\includegraphics[width=6in]{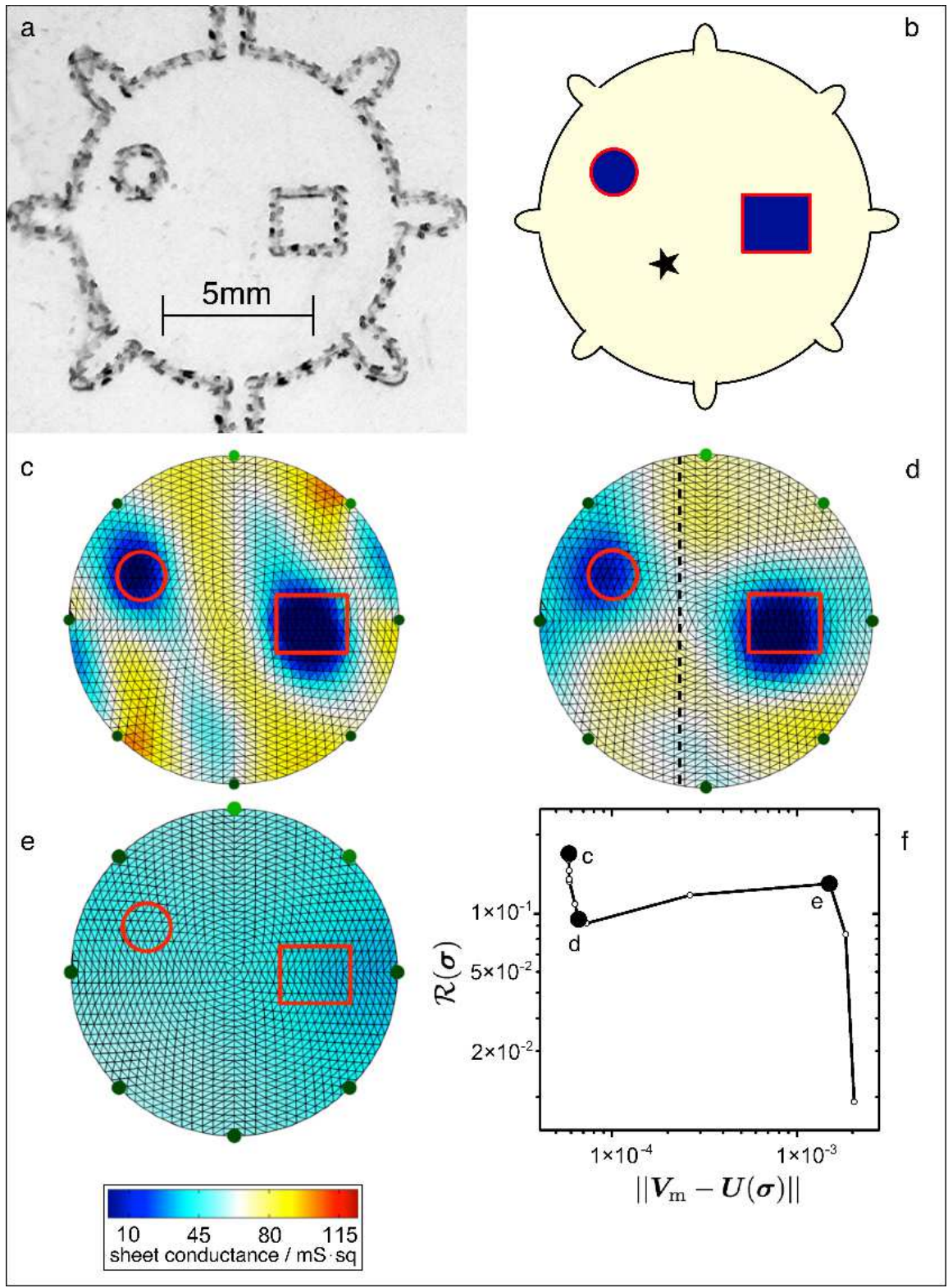}
\caption{FTOD sample. (a) photographic image. (b) schematic diagram: the coulour corresponds to the nominal sheet conductance; the marker $\star$ indicates the four-point probe sheet conductance measurement spot. (c,d,e) sheet conductance maps reconstructed with hyper-parameter values of  $8\cdot10^{-7}$ and $7\cdot10^{-5}$, $5\cdot10^{-3}$ respectively. (d) the dashed line in indicates the elements used to extract the ERT sheet conductance value of the uniform part of the sample (see table~\ref{table:results}). (f) the L-curve, with markers corresponding to the three different values of $\alpha$. The schematic diagram and the maps follow the same color-scale provided at the bottom.}
\label{fig:large}
\end{figure*}

\subsubsection{Reliability}
\begin{figure*}
\centering
\includegraphics[width=6in]{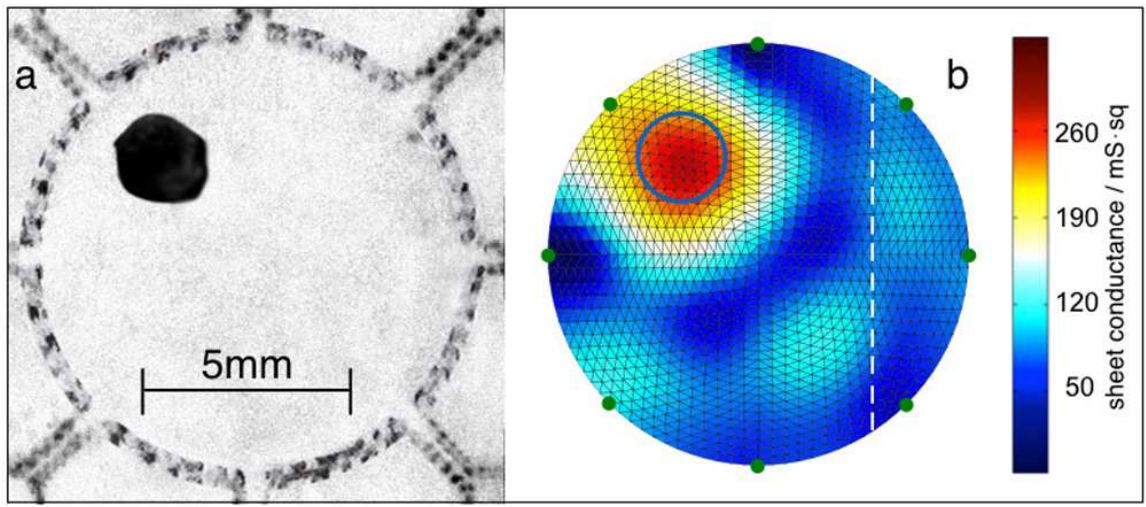}
\caption{FTOH sample. (a) photographic image. (b) sheet conductance map based on an optimal hyper-parameter value of $1.4\cdot10^{-6}$. Average sheet conductance (see table~\ref{table:results}) of the uniform zone calculated along the dashed line. }
\label{fig:ftoAg}
\end{figure*}

Fig.~\ref{fig:fto2u}-c to -e and~\ref{fig:large}-c to -e show the evolution of the reconstructions, for samples FTOU and FTOD respectively, with increasing $\alpha$ along the corresponding L-curves. 
In case of FTOU the choice of different regularization parameter did not lead to huge deviations. What is remarkable is that even for such a ``flat'' situation the corresponding L-curve features a well defined corner. Hence L-curve allowed to choose an optimal hyper-parameter value among many apparently equivalent ones. This means that the map in Fig.~\ref{fig:fto2u}-d should be assumed as the more informative given the employed model and method.
The effect of the choice of $\alpha$ on FTOD sheet conductance maps is more apparent than for FTOU (compare Fig.~\ref{fig:fto2u} and Fig.~\ref{fig:large}). A weak regularization (smaller $\alpha$) introduced some artefacts, while an excessive regularization (larger $\alpha$) cleared out any feature. On the contrary the optimal $\alpha$ produces the reconstruction which both accounts for the defects (position and value) and preserves as much as possible the uniformity far from damaged zones.
The same procedure was applied to sample FTOH with similar results and produced the optimal map in Fig.~\ref{fig:ftoAg}-b. In this case the underestimation of the numerical value of the sheet conductance of the defect is due to the effect of the PSF over a defect with very different conductance than the background. Nevertheless since thin films are usually produced to be as uniform as possible, this result confirms the versatility of this technique in this field.
\section{Conclusions}
In this paper, Electrical Resistance Tomography was successfully applied to the imaging of the sheet conductance of conductive thin film samples, either uniform and \extended{containing lower or higher conductivity defects}. The measurement setup, of very simple conception, is based on single resistance measurement channel and a switching relay matrix. Results show that ERT allows to measure both the conductivity of the continuous parts of a sample, and the position and conductivity of the defective sample zones. The image resolution is limited by the number of electrodes on the sample. The results are quantitatively confirmed by van der Pauw and four-point probe sheet conductance measurements on the same samples. The technique is not limited by the sample size, geometry or conductance magnitudes investigated in the reported case examples, opening potential applications in different areas of thin film technology.

\section{Acknowledgements}
The authors thank their colleague Danilo Sarazio for help in the realization of the samples.

\newpage

\bibliography{20160605_IEEE_IM-15-11401R_arXiv}

\end{document}